\documentclass[preprint,showpacs,preprintnumbers,amsmath,amssymb]{revtex4}
\usepackage{graphicx}
\usepackage{bm}
\usepackage{subfigure}
\begin{document}
\title{
Statistical properties of periodic orbits in 4-disk billiard system: pruning-proof property
}
\author{
Takeshi Asamizuya
\footnote[1]{asamizu@r.phys.nagoya-u.ac.jp}
}
\affiliation{Department of Physics, Faculty of Science, Nagoya University, 464-8602, Nagoya, Japan}
\date{August 11, 2005}
\begin{abstract}
%
%
Periodic orbit theory for classical hyperbolic system is very significant matter of how we can interpret spectral statistics in terms of semiclassical theory.
Although pruning is significant and generic property for almost all hyperbolic systems,
pruning-proof property for the correlation among the periodic orbits which gains a resurgence of second term of the random matrix form factor remains open problem.
In the light of the semiclassical form factor, our attention is paid to statistics for the pairs of periodic orbits.
Also in the context of pruning, we investigated statistical properties of the ``actual'' periodic orbits in 4-disk billiard system.
This analysis presents some universality for pair-orbits' statistics.
That is, even if the pruning progresses, there remains the periodic peak structure in the statistics for periodic orbit pairs.
From that property, we claim that if the periodic peak structure contributes to the correlation, namely the off-diagonal part of the semiclassical form factor,
then the correlation must remain while pruning progresse. 
\end{abstract}
\maketitle
\section{Introduction}
\subsection{Chaotic property of classical systems and universality of spectral statistics}
%
Statistical properties of periodic orbits is one of fundamental problems in quantum chaos, i.e. quantum-classical correspondence for classically chaotic systems.
Quantum chaos has been pursued in the last few decades \cite{ChaoticBehaviour1983,QCYukawa1994,CasatiChirikovQC1995,BrackBaduriSclPhys1997,SUSYTraceFormula1999}.
Especially in the context of semiclassical theory, the spectral statistics has been often referred to as an emblem of quantum-classical correspondence \cite{Richter2000,Haake2001}.
According to the BGS ( Bohigas-Giannoni-Schmit ) conjecture \cite{BGS1984}, 
there must be a universality of level statistics which goes along random matrix theory,
that is, if the classical system has stochasticity then the spectral statistics for the corresponding quantum system obeys GOE one.
But there is no self-evidence for illustrating the above facts.
%
%
\subsection{Periodic orbit theory and semiclassical analysis for spectral statistics}
%
%
Quantum - classical correspondence have been discussed mainly in spectral statistics for the last few decades.
One of the objects for exploring the correspondence is density of states,
\begin{equation}
d \left( E \right) \equiv \sum_{n} \delta \left( E - E_n \right) =
\langle d \left( E \right) \rangle  + d_{osc} \left( E \right) ,
\label{densityofstates}
\end{equation}
where $E$ is a energy and $ d \left( E \right) $, $\langle d \left( E \right) \rangle$ and $d_{osc} \left( E \right)$ are the density of states, the mean part and the fluctuating part, respectively.
%
%

%
Before the terminology "Quantum Chaos" \cite{ChaosQuantumPhysics1989,Stoeckmann1999} became used, 
Gutzwiller derived the trace formula which was prologue to proceeding of periodic orbit analysis 
\cite{Gutzwiller1990,Gutzwiller1967,Gutzwiller1969,Gutzwiller1970,Gutzwiller1971,Gutzwiller1973,Gutzwiller1977}.
\begin{equation}
d_{osc} \left( E \right) =
\frac{1}{\pi \hbar} \Re \sum_{\gamma,\kappa}
B_{\gamma} \frac{T_{\gamma}}{\kappa} 
\exp  \left[ \frac{i}{\hbar} S_{\gamma} \left( E \right) \right]
,
\label{gtf}
\end{equation}
The trace formula represents fluctuating part $d_{osc} \left( E \right)$ of spectral density of states and it is derived by semiclassical evaluation of Feynman path integrals.
$\gamma$, $S_{\gamma}$, $T_{\gamma}$, $B_{\gamma}$, $\kappa$ are index, action, period, amplitude factor and repetition number of periodic orbits, respectively.
The amplitude factor $B_{\gamma}$ consists of the Monodromy matrix $ M_{\gamma} $ and the  Maslov index $ \sigma_{\gamma} $,
\begin{equation}
B_{\gamma} = \frac{1}{\sqrt{| M_{\gamma} - I |}} \exp \left( i \frac{\pi}{2} \sigma_{\gamma} \right)
.
\label{amplitude}
\end{equation}
Admirably, the quantities which are factored in the trace formula are canonical invariant.
%
%

One of the main objects of semiclassical analysis for spectral statistics is spectral form factor,
\begin{equation}
K \left( \tau \right) = \int^{\infty}_{-\infty} \frac{d \eta}{ \langle d \left( E \right) \rangle}  
\Big\langle
d_{osc} \left( E + \frac{\eta}{2} \right) d_{osc} \left( E - \frac{\eta}{2} \right)
\Big\rangle_E
\exp \left[ - i 2 \pi \langle d \left( E \right) \rangle  \eta \tau \right]
,
\label{sff}
\end{equation}
which was proposed by Berry \cite{Berry1985}, and which is associated with semiclassical sum rule, i.e. bootstrap
\begin{equation}
\langle d \left( E \right) \rangle =
\lim_{\varepsilon \rightarrow 0} \sqrt{2} \, \varepsilon \langle d_{osc,\varepsilon}^{2} \left( E \right) \rangle
.
\label{bootstrap}
\end{equation}
On the one hand, from the random matrices theory \cite{Mehta1991,Bohigas1989}, 
the spectral form factor is obtained by applying Fourier transformation to the two points level correlation.
\begin{equation}
K^{GOE} \left( \tau \right)  = 2 \tau - \tau \ln \left( 1 + 2 \tau \right) = 2 \tau - 2 \tau^2 + \dots, \qquad \textrm{for} \quad  \tau < 1
,
\label{sffrmt}
\end{equation}
provided $\tau = T/T_{H}$, $T_{H}=2 \pi \hbar \bar{d}\left( E \right) $ is Heisenberg time.
This is valid for the systems with time reversal symmetry.
On the other hand, from semiclassical analysis, i.e. substituting the Gutzwiller's trace formula for the spectral form factor \eqref{sff}, it becomes
\begin{equation}
K \left( T \right) \approx 
\frac{1}{2 \pi \hbar \langle d \left( E \right) \rangle } 
\sum_{\gamma,\gamma^{\prime},\kappa,\kappa^{\prime}}
\Big\langle
B_{\gamma} B_{\gamma^{\prime}}^{\ast} \frac{T_{\gamma}}{\kappa} \frac{T_{\gamma^{\prime}}}{\kappa^{\prime}} 
\exp \left[ 
\frac{i}{\hbar} \left( S_{\gamma} \left( E \right) -  S_{\gamma^{\prime}} \left( E \right) \right) 
\right] 
\delta \left( T - \frac{ T_{\gamma} + T_{\gamma^{\prime}} }{2} \right)
\Big\rangle_{E,T}
,
\label{sffscl}
\end{equation}
where $\langle \cdot \rangle_{E,T}$ means smoothing over energy $E$ and time $T$.
Employing Hannay - Ozorio de Almeida sum rule ( the principle of uniformity ), Berry estimated diagonal approximation of the semiclassical form factor \eqref{sffscl}.
Then, for a chaotic system with time reversal symmetry, the diagonal approximation of the expression \eqref{sffscl} becomes
\begin{equation}
K \left( \tau \right) \sim 2 \tau \quad \textrm{for} \quad \tau < 1
.
\label{sffrmt1st}
\end{equation}
Hannay - Ozorio de Almeida sum rule \cite{Almeida1988,HO1984} represents bearing of periodic orbits' sum in hyperbolic system.
\begin{equation}
T \sum_{\gamma} \frac{1}{| M_{\gamma} - I |} \rightarrow 1, \qquad  \textrm{for} \quad T \rightarrow \infty.
\label{hodasumrule}
\end{equation}
That is, in the context of the semiclassical theory \eqref{sffscl}, 
the counterpart of the first term of the random matrix form factor \eqref{sffrmt} is interpreted as a result from the manner of the distribution of the periodic orbits.
\subsection{Progress in analysis for correlation among classical orbits}
%
%
Contrary to the achievement above mentioned, there is the significant matter which had been left unanswered; 
in the semiclassical manner, as what we can take a counterpart of the reminder term of the random matrix form factor?
As is mentioned in the preceding subsection, through the diagonal approximation of the semiclassical form factor, 
the first term of the random matrix form factor had been perceived as a result from the distributional manner of the periodic orbits.
However, the remainder terms of the random matrix form factor, $-2\tau^{2} + \cdots $ or $-\tau \log \left( 1 + 2 \tau \right)$, have not been barely accounted for by periodic orbits 
more than the fact that they can be described with the off-diagonal terms of the semiclassical form factor, i.e. combinations among classical periodic orbits except for its own combinations.
Although it becomes clear that much more amount of off-diagonal terms than the diagonal terms reduce to the corrective effect through the cancelation in phases, we don't know how it works.
In other words, the outstanding issue is how to explain  the contribution which remain through the cancelation among the phases, 
especially in the context of classical dynamics.
In other words, we have not known yet which combinations of periodic orbits make principal contribution to the off-diagonal part of the semiclassical form factor 
which corresponds with the remainder terms of the random matrix form factor.

Toward understanding the relation between the off-diagonal terms of the semiclassical from factor \eqref{sffscl} and the remainder terms of random matrix form factor \eqref{sffrmt},
many endeavours have been made \cite{ADDKKSS1993,AIS1993,BK1996,Sano2000}.
When it became the 21st century, Sieber and Richter \cite{SR2001} presented an approach:
they assumed the pair of periodic orbits which one orbit had self-crossing and another enclosed one and the certain distribution which the periodic orbits obeyed, 
and using semiclassical analysis, they derived the counterpart of the reminder term of the random matrix form factor \eqref{sffrmt}.
This pioneering approach and those following works share some ideas; the logarithm dependency of the target distribution, the uniqueness of partner orbits, etc...;
and they prospered in derivation of second term of the form factor analytically by use of semiclassical theory 
\cite{Sieber2002,BHH2002,Muller2003,TR2003,Spehner2003,HMBH2004,MHBHA2004,TSMR2005}.

Apart from the direct analysis for the spectral form factor, 
Shudo and Ikeda \cite{SI1994} investigated action pair difference in the kicked rotor system 
and they claimed significancy of the periodic peak structure of the action pair distribution in the context of dynamical localisation.
Concretely saying, they presented the classification of the pairs of (non-periodic) orbits which contribute to the periodic peak structure with their symbol sequences: 
the orbits pairs have the symbol sequences which have period-4 cycle structure and have some other symbols inserted in the cycle structure.
And they concluded that such classified orbits were the homoclinic orbits.

Other approaches on correlation among periodic orbits are also interesting.
For example, Efetof and Kogan \cite{EK2003} derived the results of the random matrices theory via the non-linear sigma model in disordered electronic systems.
The diagonal approximation of their results is equivalent to the lowest order perturbation theory  ( Altshuler and Shklovskii \cite{AS1986} ).
and the higher order corrections would be equivalent to correlation which consists of pairs among periodic orbits.
\subsection{What is the object and what is the condition to be required for the periodic orbits' correlation?}
%
As just described, many works have been performed in this decade, but with all those works, it is not enough to understand the correlation.
There remain several questions and rack of information to understand what is the correlation; 
it is not clear whether there is consistency between the periodic peak structure in the Shudo and Ikeda's work and the Sieber and Richter's idea; 
there is few researches to show whether or how "pruning" phenomena have an effect on the correlation; 
some researches posed the distribution of the periodic orbits which can contribute to the semiclassical form factor, 
but it is not clear that the other periodic orbits not to contribute to the semiclassical form factor can be vanished into nothing through cancelation among phases; 
there is a few works using actual numerical periodic orbits, but it is not sufficient to understand the correlation;
if there is the correlation, 
then it becomes recognisable in statistics of not only implicit physical quantities, e.g. crossings of periodic orbits, but also obvious physical quantities, e.g. their actions and periods, 
but there is little direct analysis of the statistics for the obvious quantities.

In accordance with the lessons above mentioned, we will discuss the object and the conditions for a correlation among periodic orbits as a rule of thumb.
Requiring such conditions, we will find out the correlation in classically hyperbolic system.
As already mentioned, statistical property of periodic orbits is absolutely essential for understanding spectral statistics in semiclassical manner.
However, it is too hard to solve the above problems simultaneously.
Hence, we must winepress into the following two points; 
(a) what is the variable for which the correlation must be appeared in the statistics?; 
(b) what is the condition which the correlation must satisfy?

\begin{description}
%
\item[(a) what is the variable for which the correlation must be found in?:]@\\
The statistics or the correlation to be observed must be measured in terms of the quantities which the semiclassical form factor contains explicitly,
e.g. action $S$, period $T$, Monodromy matrix $M$, etc... .
Otherwise, as a future task, it is hard to understand how the discovered correlation contribute to the semiclassical form factor and recreate the remainder terms of the random matrix form factor.
From the viewpoint of explicitly, for example, to treat crossing angle as the object for semiclassical analysis has disadvantage, 
because crossing angle is not contained explicitly in the semiclassical form factor and it is not canonically invariant and also it cannot be defined in some systems.
And what is more, we believe that if the correlation which contributes to the semiclassical form factor does exist, 
then they must emerge also in an obvious expression, e.g. statistics among actions and periods for periodic orbits.
Therefore, if we take a statistics about the semiclassical form factor among several variables, then we will give the highest priority to the action and the period.
Because from the representation of the semiclassical form factor, it is clear that the most important quantities are not the stability and the Maslov index but the action and the period.
Telling as it is, in the semiclassical form factor, there are the differences between the actions in the phases and the sums of the period in the delta functions.
It is legitimate to consider that cancelation among phases and delta functions play a significant role for contributing to the semiclassical form factor.
%
\item[(b) what is the condition which the correlation must satisfy?:]@\\ 
We impose ``pruning-proof'' property on the correlation what we must look for.
In the context of ``pruning'' \cite{Cvitanovic1991}, the correlation had not been thoroughly understood.
Pruning is a phenomenon that structurally unstable periodic orbits are disappeared as a system parameter is varying.
And pruning is a general property for any hyperbolic systems \cite{Ishii1997N,Ishii1997CMP,HansenPHD,Hansen1993a,Hansen1993b,HS2004a,HS2004b}.
The terminology ``pruning'' means, plainly speaking, breakdown of 1-to-1 correspondence between periodic orbits and symbol sequences, 
that is, number of periodic orbits decreases from how much ones ``complete system'' should have essentially ( cf. subsections II B and C ).
The word "pruning-proof" property represents an immune property against such pruning process.
It is an important matter whether such the correlation is independent of pruning.
If a discovered correlation has not pruning-proof property, then such correlation cannot be called as universal property any longer, 
and then it is inadequate to adopt that correlation as significant contribution to the semiclassical form factor.
Therefore, it is indispensable to examine whether a discovered correlation has pruning-proof property. 
If a correlation among periodic orbits constructs significant contribution to the semiclassical form factor, then the correlation must have a pruning-proof property.
That is, pruning-proof property is a necessary condition for the correlation which gives some contribution to the semiclassical form factor.
\end{description}
\subsection{Aims of this paper}
%
%
The aims of this work are the following two points; 
(a) to find out a correlation among periodic orbits which must be measured in terms of the quantities of which the semiclassical form factor is apparently consists;
(b) to confirm that the observed correlation is proof against pruning.
If the above aims are achieved, then it turns out that we get a candidate for the correlation which contribute to the semiclassical form factor.

%
In addition, we must mention that we ease the condition along which a correlation among periodic orbits must go.
Meanwhile, we don't mind whether the correlation recurrence the form of the remainder terms of the random matrix form factor.
Actually, with a numerical method, 
it is too difficult to recreate the equivalent of the remainder terms of the random matrix form factor with the correlation as well as only to find out the correlation itself.
And then we intend to postpone understanding how a periodic orbits' correlation contributes to the semiclassical form factor in this paper.
Therefore we would mind only what kind of correlation is there in classically hyperbolic system.  
\subsection{Construction of this article}
In Section II, we will illustrate the objective system, specifically 4-disk billiard.
In Section III, we will state the methods of analysis, especially the pair statistics for periodic orbits.
In Section III and IV, we will state the statistics for the single periodic orbits to support the pair statistics.
In Section V, we will present the results for the above pairs' and singles' analysis.
In Section VI, we will give summary and discuss the relation between the correlation and the pruning.
\section{system}
\subsection{4-disk billiard}
\begin{figure}
\begin{center}
\hspace{0cm}
\subfigure[]{\includegraphics[width=3.8cm,clip]{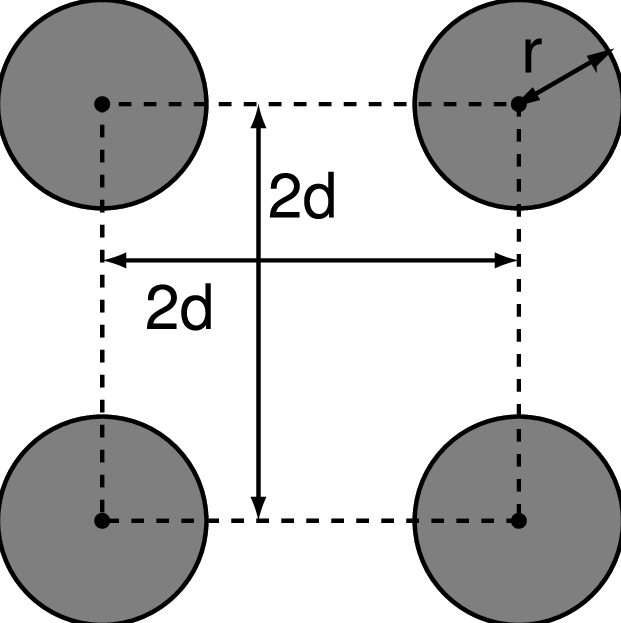}}
\hspace{0mm}
\subfigure[]{\includegraphics[width=3.8cm]{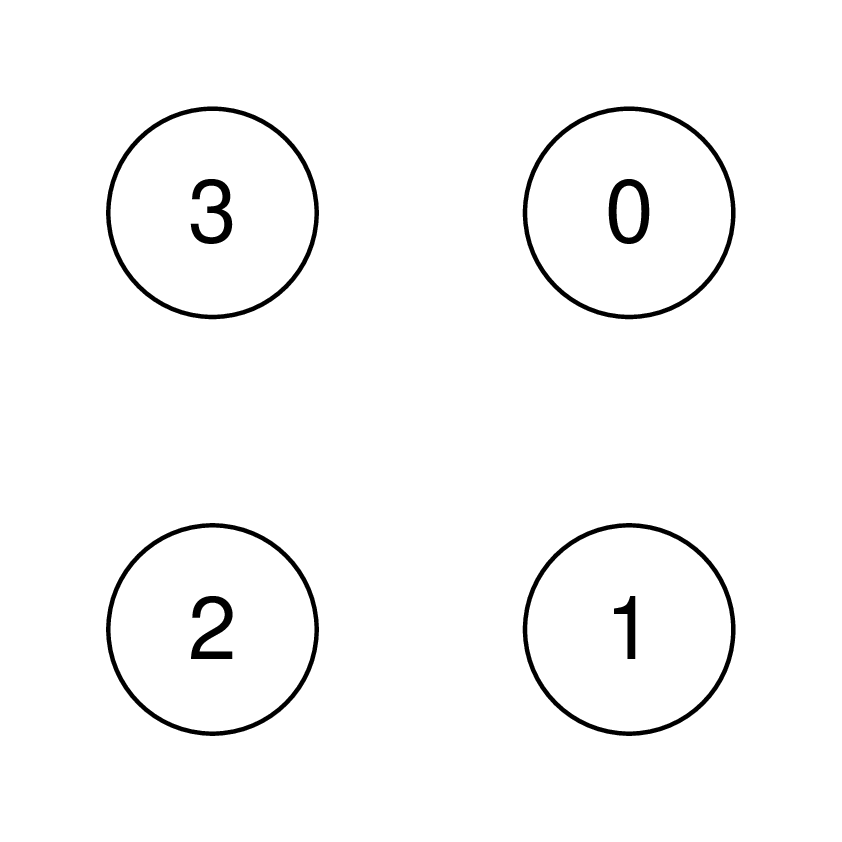}}
\hspace{0mm}
\subfigure[]{\includegraphics[width=3.8cm,clip]{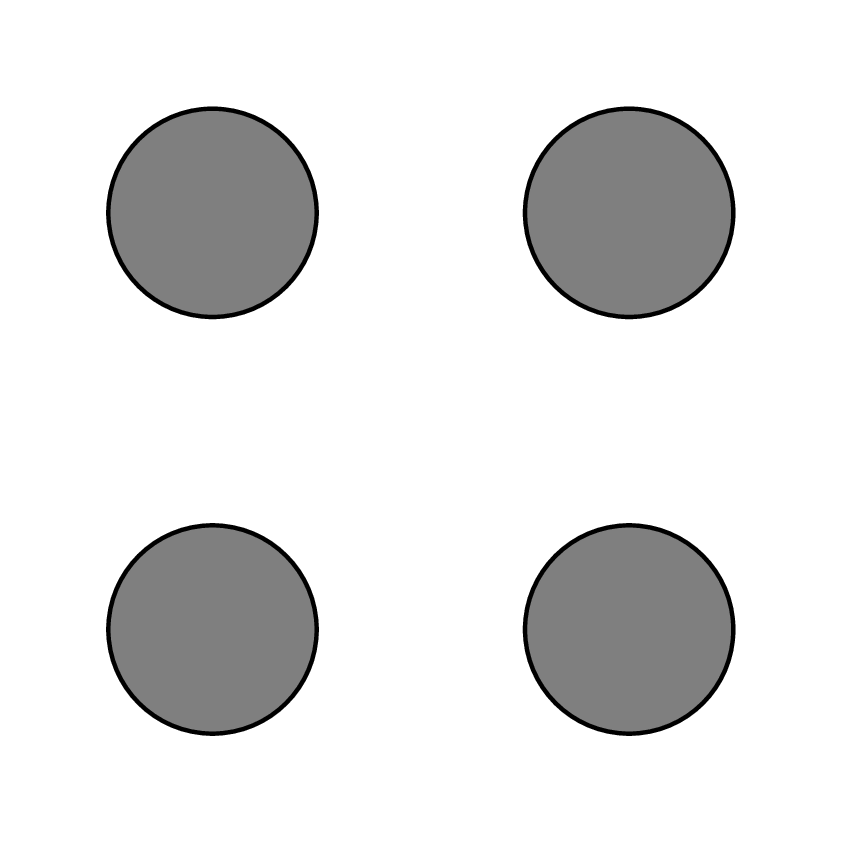}}
\hspace{0mm}
\subfigure[]{\includegraphics[width=3.8cm,clip]{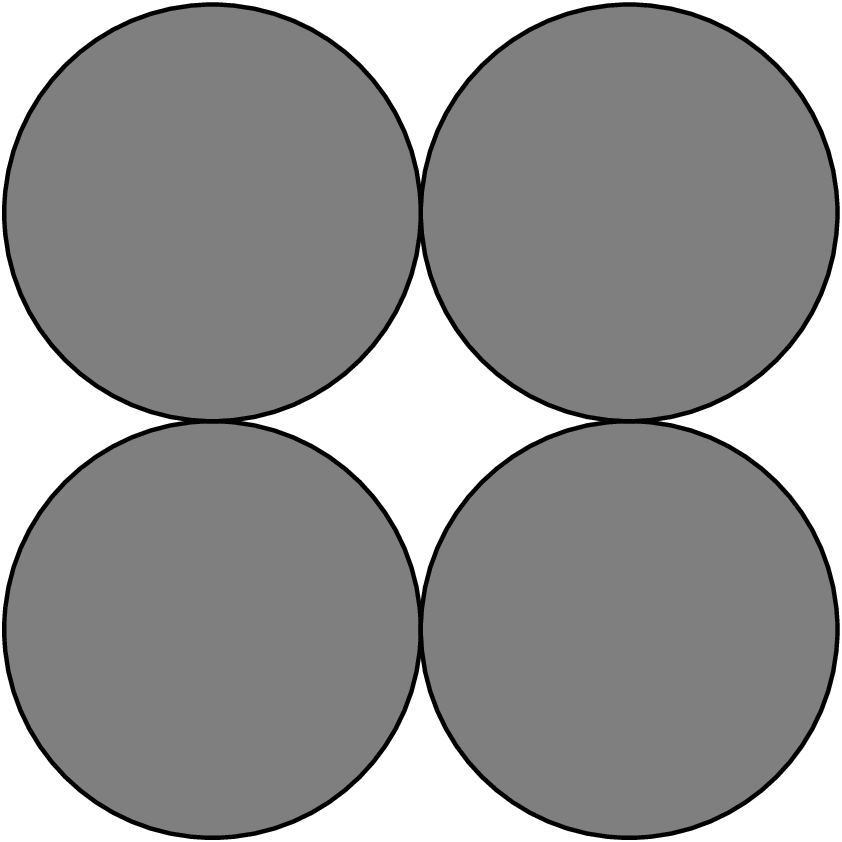}}
\caption{{\bf 4-disk billiards} : (a) parameter setting, (b) the numbered 4 disks, (c) when $r=d/2=0.5$, (c) when $r=d=1.0$}
\label{4diskbilliard}
\end{center}
\end{figure}
Here we illustrate the object system, 4-disk billiard system \cite{Gaspard1998}.
Billiard system is a model for idealised motion of a billiard ball ( point particle ) \cite{HardBallSysLorenGas2002}
which goes straight ahead in a given domain and collides elastically on a boundary.
First, the definition of this system is given for the configuration of the domain $\mathcal{Q}$ in which billiard ball runs ( Fig.\ref{4diskbilliard} ):
\begin{eqnarray}
\mathcal{Q}
 \in \mathbf{R}^{2} 
\setminus
\mathcal{D}
,\quad
\mathcal{D}= \mathcal{D}_{0} \cup \mathcal{D}_{1} \cup \mathcal{D}_{2} \cup \mathcal{D}_{3},
\\
\mathcal{D}_{i}
\equiv
 \{ 
\left( x, y \right) \in \mathbf{R}^{2} 
\bigm| 
 \left( x - x_{ci} \right)^2 + \left( y - y_{ci} \right)^2 < r^2
\},
\quad \textrm{for} \quad  i = 0,1,2,3\,,
\label{defbiliard}
\end{eqnarray}
where the radius $r$ is common to all the four disks and $2d$ is the side of the square that each centres of disks is located on the four vertices respectively.
Let us name each of the four disks as "$i$-th disk" in clockwise order from the top right disk ( Fig.\ref{4diskbilliard} (b) ).
Each $i$-th disk is arranged its centre on the four vertices of the square respectively, that is, 
$\left( x_{c0}, y_{c0} \right) = \left( d, d \right)$, $\left( x_{c1}, y_{c1} \right) = \left( d, -d \right)$, 
$\left( x_{c2}, y_{c2} \right) = \left( -d, -d \right)$, $\left( x_{c3}, y_{c3} \right) = \left( -d, d \right)$.
$\mathcal{D}_{i}$ is the interior of each $i$-th disk and $\mathcal{D}$ is the total interior area of all the four disks.
The domain $\mathcal{Q}$ is defined as the complement of $\mathbf{R}^2$ with respect to $\mathcal{D}$.
The billiard ball goes freely within the domain $\mathcal{Q}$ and collides elastically on the boundary $\partial \mathcal{Q}$.
When $r=1.0$, the four disks touches to each other, and when $r>1.0$, the four disks overlap each other.
In the case of $r \geq 1.0$, the four disks divides the 2-dimensional configuration space into the two domains, 
one is enclosed by them and another encloses them.
In any case, the periodic orbits can be in the area enclosed by 4-disks, $ \left( -d \leq x,y, \leq d \right) \cup \mathcal{Q}$.

In this system, it is easy to calculate periodic orbits 
in the sense of adequate correspondence between periodic orbits and symbol sequences \cite{Bunimovich1995},\cite{Morita1991}.
For example, the billiard ball bounce with a succession of the $i$-th disks, in such way, ``0'', ``1'' , ``2'', ``3'', 
and then corresponding symbol sequence is ``0123'' ( see Fig.\ref{4diskbilliard} and Fig.\ref{pruningiebifurcation} ).
\subsection{Pruning}
\begin{figure}[htbp]
\begin{center}
\includegraphics[width=5cm]{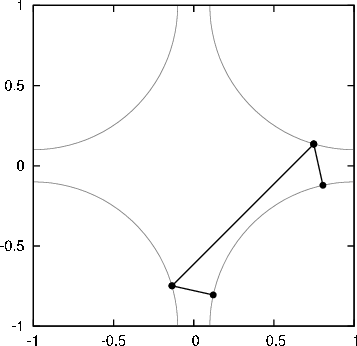}
\includegraphics[width=5cm]{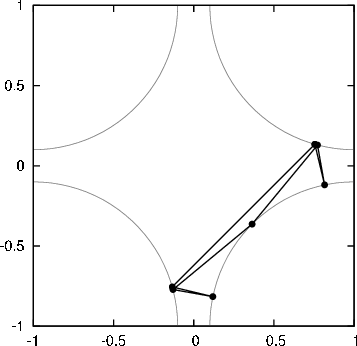}
\includegraphics[width=5cm]{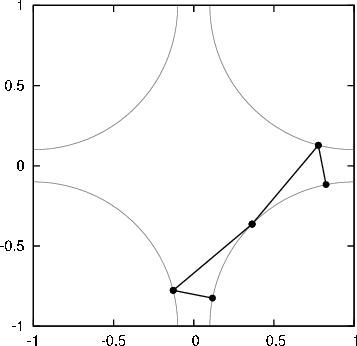}
\\
\includegraphics[width=5cm]{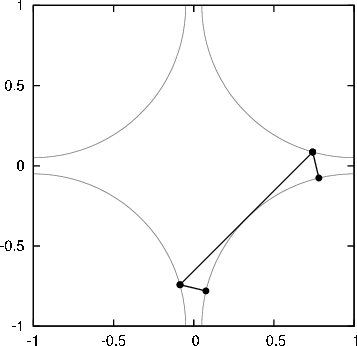}
\includegraphics[width=5cm]{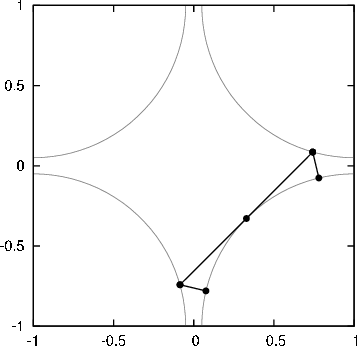}
\includegraphics[width=5cm]{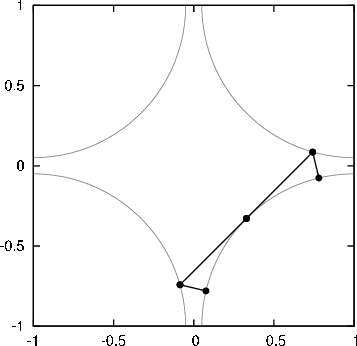}
\\
\includegraphics[width=5cm]{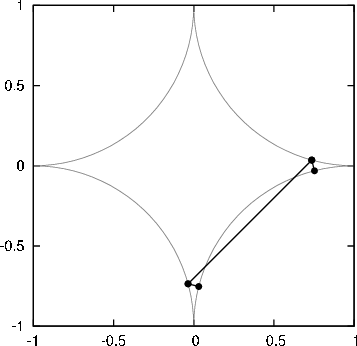}
\includegraphics[width=5cm]{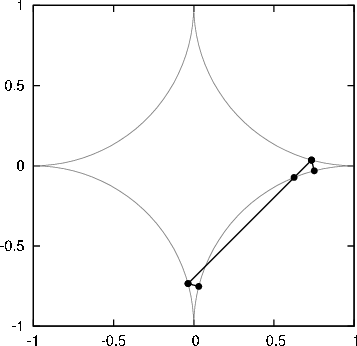}
\includegraphics[width=5cm]{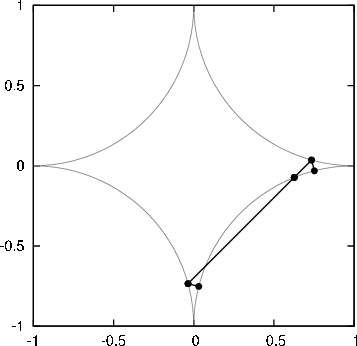}
\caption{ {\bf To be pruned, i.e. to be bifurcated, orbits in 4-disk billiard}:
from left to right, orbit has '010212'/ '0101212' / '01012121' symbol sequence and has 6 / 7 / 8  collisions.
from top to bottom, $r=0.9,\,0.95,\,1.0$.
They become pruned with going to downward in these figures, while they become bifurcated with going to upward.
}
\label{pruningiebifurcation}
\end{center}
\end{figure}

%
%
Pruning is, as its name suggests, trimming in symbol sequence's space. 
Primarily, in generally hyperbolic systems, any periodic orbit can be transliterate into an appropriate symbol sequence,
but it is not necessarily vice versa.
That is, it is not always true that there is periodic orbit corresponding to any symbol sequence.
We describe this situation as ``pruned'' state or state that system has pruning property.
In other words, pruned state is that there is physically impossible periodic orbit which is hypostatised from certain symbol sequence ( Fig.\ref{pruningiebifurcation} ).
To put in another way, if there is a complete symbol sequence's space,
we can regard it as a ``tree'' and call a symbol sequence as a ``branch'' of the tree.
Pruning is defection of appropriate branches among the tree.
\subsection{Pruning in 4-disk billiard system}
%
%
Let us review the condition and the circumstance of pruning for the 4-disk billiards.
All the four disks have the same radius $r$ which is the system parameter to be varied from $r=0.5$ to $r=1.0$ simultaneously.
The fact to be aware of is that variable is radius $r$, not distance $2d$ , apart from many of foregoing works.

First, we consider not prunend state, i.e. corresponding symbol sequence's space is complete one.
When $r=0.5$, there is no eclipse, more precisely, looking from one disk toward diagonal one, there is no obstacle to interrupt the view.
Even on the morrow of the eclipse ($r_{e}=d/\sqrt{2}=0.70710678..$), the corresponding space of symbol sequences has no pruning.
Eventually up to $r=r_{c}=0.90715514863324558...$, pruning does not occur.

\begin{figure}[htbp]
\begin{center}
\includegraphics[width=6cm]{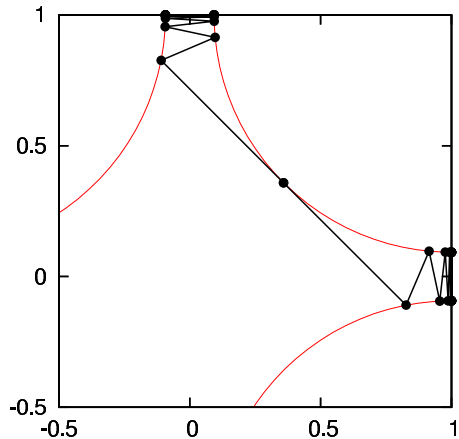}
\caption{{\bf Instant of first pruning in 4-disk billiard} ($r=r_{c}=0.90715514863324558...$)}
\label{1stpruning}
\end{center}
\end{figure}

Next, when $r=r_{c}=0.90715514863324558...$, first-pruned orbit touches on the $0$-th disk \cite{HansenPHD} ( Fig.\ref{1stpruning} ).
The periodic orbit to be pruned first among the whole set of periodic orbits has the symbol sequence, 
e.g. ``$0(30)^n31(01)^n0(10)^n13(03)^n$'' and ``$3(03)^n(10)^n1(01)^n(30)^n$'', etc. ($n\rightarrow \infty$),
that is, after infinite number of collisions between $3$rd-disk and $0$-th disk, there are also infinite ones between the $0$-th disk and the $1$-st disk and again.

%
Finally, after the fist pruning, the above first-pruned orbit becomes physically impossible.
In this way, from $r=r_{c}$, pruning occurs successively as $r$ is increased ( Fig.\ref{pruningiebifurcation} ).

In the 4-disk billiard system, pruning can be interpreted as reverse to bifurcations.
That is, reducing the radius, physically impossible orbit come to realistic one, which we call this as ``to be pruned'' orbit, and simultaneously it becomes bifurcated.
Conversely, as enlarging the radius, bifurcated orbits are integrated into a single orbit, 
and immediately such an orbit disappears as a ghost one, i.e. physically impossible orbit ( Fig.\ref{pruningiebifurcation} ).
\subsection{Numerical calculation of periodic orbits}
%
%
With the aid of not true periodic orbits but non-periodic orbits, which must be near to ``true'' one,
the studies of semiclassical analysis of the spectral form factor have been achieved.
We should not overlook that there might be ``pruning'' among the periodic orbits 
which has the same symbol sequence as that of the non-periodic orbit which is expected to contribute to form factor.
Therefore, ``actual'' periodic orbits is indispensable to semiclassical analysis with the aid of the trace formula \eqref{gtf}.

We calculated periodic orbits numerically, as the preparatory to the investigation of periodic orbits' statistics.
Actual method to calculate any periodic orbit is achieved by minimisation of whole ad referendum length of one \cite{Bunimovich1995}, \cite{CMP1987}, \cite{HS1992a}.
We perform this optimisation by use of Newton-Raphson's method.
To avoid 'chaotic behavior' of Newton-Raphson's method, we also use bi-sectional method before using the former method.
First, making use of bi-sectional method, we can see that the numerical periodic orbit converges in the neighbourhood of the ``true'' periodic orbit within adequate range.
Next, using Newton-Raphson's method, we converge numerical periodic orbits to the ``true'' ones within more narrow error range.
In this way, we get ``actual'' periodic orbits numerically.

Actually, the amount of the periodic orbits which we have calculated in 4-disk billiard systems are approximately 1,000,000 for every value $r$ of radius which we taken.
The limitation of this calculation is the frequency of collisions which billiard ball impacts against 4 disks.
We investigate through 20 time collisions.
For generating long symbol sequence which produce appropriate periodic orbit, we treated long integer with the aid of CLN ( Class Library for Numbers ) library \cite{CLN}.
We actually use periodic orbits for pairs' statistics to 17 times collision for the reason why there are both limitations, the outperform of computers and the times.
The number of the periodic orbits which has 17 times collisions is $949560$ in $r=0.5$ ( without pruning ), and $703999$ in $r=1.0$ ( with pruning ).
\section{numerical analysis I : statistics of pairs of periodic obits}
\subsection{Analysis to do}
In this section, we will define statistics to analyse for correlation among pairs of periodic orbits.
And we will state results of these numerical analysis in Section V.
As we stated above, statistics of periodic orbits, specifically statistics of pairs among them, has became very significant matter in semiclassical analysis for spectral statistics.
From consistency among Eqs. \eqref{sffscl},\eqref{sffrmt}, and \eqref{sffrmt1st},
it is clear that a statistic on analysis beyond diagonal approximation are inevitable.
Therefore we will investigate statistical properties of periodic orbits, more precisely, of pairs among them.
In the semiclassical representation of the spectral form factor \eqref{sffscl},
there are the difference in action between pair of the periodic orbits ( $\gamma, \gamma^{\prime}$ ) included in the phase factor, $S_{\gamma} - S_{\gamma^{\prime}}$,
and the sum in period between pair of the periodic orbits included in the delta function, $T_{\gamma} + T_{\gamma^{\prime}}$.
Then, in working up statistics in reference to pairs among periodic orbits,
we take particular note of not only differences between actions, $S_{\gamma} - S_{\gamma^{\prime}}$, 
but also sums of periods, $T_{\gamma} + T_{\gamma^{\prime}}$,
in distinction from the former analysis.
\subsection{The spectral form factor in billiard system}
%
%
As noted above, in the semiclassical spectral form factor \eqref{sffscl}, 
differences between actions $S_{\gamma} - S_{\gamma^{\prime}}$ and sums of periods $T_{\gamma} + T_{\gamma^{\prime}}$ play significant role.
Here we will illustrate that the matter of the correlation among periodic orbits, i.e. the contribution of the off-diagonal part toward the semiclassical form factor, 
reduce to that of the pair length statistics in a billiard system.
For this end, we begin with the spectral form factor in a billiard system without external field.
In a planer billiard system, utilising length of periodic orbit $L_{\gamma}$, action and period for any periodic orbit become the following:
\begin{eqnarray}
\nonumber
S_{\gamma} \left( E \right) &=& \sqrt{2 m E} L_{\gamma} \rightarrow \sqrt{2 E} L_{\gamma}, \\ \nonumber
T_{\gamma} \left( E \right) &=& \frac{L_{\gamma}}{\sqrt{\frac{2 E}{m}}} \rightarrow \frac{L_{\gamma}}{\sqrt{2 E}}, 
\label{periodactionforbilliard}
\end{eqnarray}
where we substitute unity for mass of a billiard ball, $m=1$.
Then the semiclassical form factor is represented as
\begin{equation}
K \left( L \right) \approx 
\frac{1}{ T_{H}} 
\sum_{\gamma,\gamma^{\prime},\kappa,\kappa^{\prime}}
\Big\langle
B_{\gamma} B_{\gamma^{\prime}}^{\ast}  \frac{ L_{\gamma} L_{\gamma^{\prime}} }{ \kappa \kappa^{\prime}} \frac{1}{\sqrt{2 E}}  
\exp \left[ 
\frac{i}{\hbar} \sqrt{2 E} \left( L_{\gamma} -  L_{\gamma^{\prime}} \right) 
\right] 
\delta \left( L - \frac{ L_{\gamma} + L_{\gamma^{\prime}} }{2} \right)
\Big\rangle_{E,L} 
,
\label{sffbilliard}
\end{equation}
that is, the spectral form factor becomes a function of periodic orbit's length $L$.
Eventually question as to action difference and period sum are rehashed into that as to length difference and length sum.
Therefore, statistics on length difference and length sum over periodic orbit is indispensable to consider a candidate for the correlation contributing to the semiclassical form factor.
\subsection{Distribution function of length difference / sum for periodic orbit pairs $C^{\pm}_{\Delta L_{\pm},m} \left( L_{\pm} \right)$}
%
%
For the reasons which have been mentioned above, we define the statistics with regard to the pair statistics for periodic orbits.
From brief consideration, the most fundamental statistics for pairs of the periodic orbits are distribution of length difference or sum for every two periodic orbits.

Here we define the following statistics:
\begin{eqnarray}
C^{-}_{\Delta L_{-}, m} \left( L_{-} \right) &\equiv&
\sharp \{ \gamma \neq \gamma^{\prime} \bigm| 
m \Delta L_{-} \leq L_{\gamma} - L_{\gamma^{\prime}}  < \left( m +1 \right) \Delta L_{-} \} ,
\label{lpd}
\\
C^{+}_{\Delta L_{+},m} \left( L_{+} \right) &\equiv&
\sharp \{ \gamma \neq \gamma^{\prime} \bigm| 
m \Delta L_{+} \leq  L_{\gamma} + L_{\gamma^{\prime}}  < \left( m +1 \right) \Delta L_{+} \} ,
\label{lps}
\end{eqnarray}
\begin{equation}
L_{\pm} = L_{\gamma} \pm L_{\gamma^{\prime}},
\quad
\Delta L_{\pm} = \sup_{\gamma} \{ L_{\gamma} \pm L_{\gamma^{\prime}} \} /M,
\quad
m = 0, 1,  \dots ,  M-1 \in \mathbf{N}.
\end{equation}
$C^{-}_{\Delta L_{-},m} \left( L_{-} \right)$ / $C^{+}_{\Delta L_{+},m} \left( L_{+} \right)$ is the distribution function of length difference / sum for periodic orbit pairs.
The distribution function of length difference / sum is the histogram 
which represents how many pairs of periodic orbits with their length difference / sum $L_{\pm}$, are there in each divided segments,
$[ m \Delta L_{\pm},\, \left( m +1 \right) \Delta L_{\pm} ]$.
The domain of the histograms is from $0$ to supremum of $L_{\gamma} \pm L_{\gamma^{\prime}}$ among given periodic orbits, 
i.e. $[0,\,\sup_{\gamma} \{ L_{\gamma} \pm L_{\gamma^{\prime}} \} ]$.
$M$ is the number of bins which divides the domain of the distribution function, and  $m$ is a index to point out a bin.
$ \gamma, \gamma^{\prime} $ are the indices of periodic orbits which includes repeated ones.
$\Delta L_{-}$ / $\Delta L_{+}$ is a class interval which classify any periodic orbits pair according to their pair length $L_{\pm}$
and they are each derived by dividing the supremum of the length difference / sum among the periodic orbit pairs into the number of bins.
In the remainder of this paper, we abbreviate $C^{-}_{\Delta L_{-},m} \left( L_{-} \right)$ / $C^{\pm}_{\Delta L_{+},m} \left( L_{+} \right)$ 
as $C^{-} \left( L_{-} \right)$ / $C^{+} \left( L_{+} \right)$ respectively.

Actually, we take the distribution $C^{-} \left( L_{-} \right)$ / $C^{+} \left( L_{+} \right)$ over the all periodic orbits with $n=17$ collisions 
and we take the distribution for each the radius of the four disks $r$, from $0.5$ to $1.0$ in the matter of 0.05.
Each of the domains $[0,\,\sup_{\gamma} \{ L_{\gamma} \pm L_{\gamma^{\prime}} \} ]$ is segmentalised into $M=2^{19}$ intervals and these distributions are not normalised.
\subsection{Joint distribution function of length difference \textit{and} sum for periodic orbit pairs $C^{\pm}_{\Delta L_{\pm},m_{-},m_{+}} \left( L_{-}, L_{+} \right)$}
%
%
In addition to the above statistics, there is need for more adequate analysis which consider length difference and its sum simultaneously.
If the correlation which contributes to the off-diagonal part of the semiclassical form factor \eqref{sffscl} is originated from the cancellation in the phase factor,
Therefore, it is needed to take statistics for periodic orbit pairs which sums of their two periods are the same value. 
In the words of \eqref{sffbilliard}, we have to investigate statistics for pairs which sums of both lengths are the same value. 

Therefore, we define the following joint distribution function :
\begin{eqnarray}
\nonumber
C^{\pm}_{\Delta L_{\pm},m_{-},m_{+}} \left( L_{-}, L_{+} \right) 
 \equiv 
\sharp \{ \gamma \neq \gamma^{\prime} \bigm| 
m_{-} \Delta L_{-} \leq L_{\gamma} - L_{\gamma^{\prime} } < \left( m_{-} +1 \right) \Delta L_{-} \, ,
\\
m_{+} \Delta L_{+} \leq L_{\gamma} + L_{\gamma^{\prime} } < \left( m_{+} +1 \right) \Delta L_{+} \}
,
\label{lpdas}
\end{eqnarray}
\begin{equation}
\nonumber
L_{\pm} = L_{\gamma} \pm L_{\gamma^{\prime}},
\quad
\Delta L_{\pm} = \sup_{\gamma} \{ L_{\gamma} \pm L_{\gamma^{\prime}} \} /M_{\pm},
\quad
m_{\pm} = 0, 1,  \dots ,  M_{\pm}-1 \in \mathbf{N}.
\end{equation}
We describe this function $C^{\pm}_{\Delta L_{\pm},m_{-},m_{+}} \left( L_{-}, L_{+} \right)$ as the joint distribution of length difference \textit{and} sum for periodic orbit pairs.
In the remainder of the paper, we abbreviate $C^{\pm}_{\Delta L_{\pm},m_{-},m_{+}} \left( L_{-}, L_{+} \right)$ as $C^{\pm} \left( L_{-}, L_{+} \right)$.
The joint distribution function, assemblage of both action differences and period sums, 
is two-dimensional histogram and represents how many pairs of periodic orbits are there in each divided cell,
$[ m_{+} \Delta L_{+}, \, \left( m_{+} +1 \right) \Delta L_{+} ] \times [ m_{-} \Delta L_{-}, \, \left( m_{-} +1 \right) \Delta L_{-} ]$.
Each of the two ranges of the variables $L_{\pm}$ of the joint distribution
is from $0$ to supremum of $L_{\gamma} \pm L_{\gamma^{\prime}}$ among periodic orbit pairs,
$[ 0, \, \sup_{\gamma} \{ L_{\gamma} \pm L_{\gamma^{\prime}} \} ]$.

\begin{figure}
\begin{center}
\hspace{0cm}
\subfigure[]{
\includegraphics[height=5.3cm,clip]{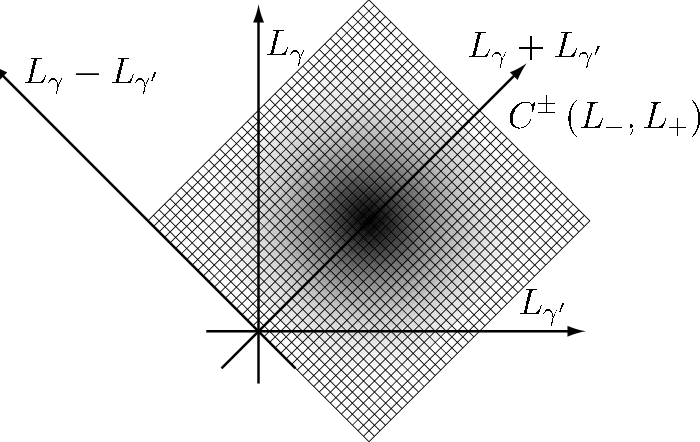}}
\hspace{0mm}
\subfigure[]{
\includegraphics[height=5.3cm,clip]{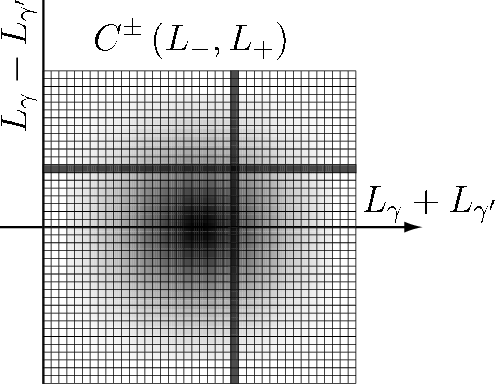}}
\caption{{\bf Joint distribution function of length difference \textit{and} sum for periodic orbit pairs:}
(a) coordinate system, (b) slicing joint distribution of length difference \textit{and} sum along vertical direction $L_{+}= \textrm{const.}$ and along horizontal direction $L_{-}= \textrm{const.}$}
\label{lpdascoordinate}
\end{center}
\end{figure}

Let us explain the other factors of this joint distribution function.
$M_{-}$ / $M_{+}$ is the number of bins which divide length difference / sum for periodic orbit pairs respectively, 
and  $m_{-}$ / $m_{+}$ is the indices of each numbered intervals $[ m_{\pm} \Delta L_{\pm}, \, \left( m_{\pm} +1 \right) \Delta L_{\pm} ]$.
$ \gamma, \gamma^{\prime} $ are the indices of periodic orbits which includes repeated ones.
$\Delta L_{\pm}$ is the class interval which categorise the pairs of periodic orbits with respect to,
and it is each defined as the supremum of the length difference / sum $\sup_{\gamma} \{ L_{\gamma} \pm L_{\gamma^{\prime}} \}$ 
among the periodic orbit pairs divided by the bin number, 
as is the case with $C^{-} \left( L_{-} \right)$ and $C^{+} \left( L_{+} \right)$ ( Eqs.\eqref{lpd}, \eqref{lps} ).
We introduce to the coordinate system ( see Fig. \ref{lpdascoordinate} ).
First, let us consider the original coordinate, $ L_{\gamma}$-$L_{\gamma^{\prime}}$, system.
If the original axes are $ L_{\gamma}$ and $ L_{\gamma^{\prime}}$,
and then the new axes are $L_{-} = L_{\gamma} - L_{\gamma^{\prime}}$ and $L_{+} = L_{\gamma} + L_{\gamma^{\prime}}$ 
which are rotated by $\frac{\pi}{4}$ ( Fig.\ref{lpdascoordinate} )
and this coordinate system is the object to be considered.

The joint distribution functions $C^{\pm} \left( L_{-}, L_{+} \right)$ are represented as a density plot.
Any density plot for logarithm of joint distribution $\log C^{\pm} \left( L_{-}, L_{+} \right)$ are expressed as a colour map.
Actually, the whole domain of the joint distribution $[ 0, \, \sup_{\gamma} \{ L_{\gamma} - L_{\gamma^{\prime}} \} ] \times [ 0, \, \sup_{\gamma} \{ L_{\gamma} + L_{\gamma^{\prime}} \} ]$
is partitioned into $M_{-} \times M_{+} = 2^{12} \times 2^{12}$ cells 
and these joint distributions are not normalised.
For each logarithm of the joint distribution $\log C^{\pm} \left( L_{-}, L_{+} \right)$, the common colour map is applied to each of the different range
which is determined according to the maximum and the minimum values among the function $\log C^{\pm} \left( L_{-}, L_{+} \right)$.
As in the case of preceding subsection, we take the joint distribution over all of the periodic orbits with $n=17$ collisions 
and we take the joint distribution for each the radius of the four disks $r$, from $0.5$ to $1.0$ in the matter of 0.05.
\subsection{Slice of joint distribution function of length difference \textit{and} sum $C^{\pm}_{\Delta L_{\pm},m_{-},m_{+}} \left( L_{-}, L_{+} \right)$}
%
%
For searching the correlation, it is insufficient only to analyse the foregoing density plots 
because in logarithm of the distribution function $\log C^{\pm} \left( L_{-}, L_{+} \right)$, the fluctuating part, which may have detailed structure, become more subtle.
Then we take the joint distribution in the form of anti-logarithm, $C^{\pm} \left( L_{-}, L_{+} \right)$.
Moreover, the most straightforward idea to analyse the joint distribution are to behold their cross-section.

For this reason, we also use following slice of the joint distribution function of length difference \textit{and} sum for periodic orbit pairs.
Briefly, the target statistics comes to the 1-dimensional histogram.
The slice of the joint distribution $C^{\pm} \left( L_{-}, L_{+} \right)$ is altogether section of it for the 2-dimensional histogram.
In equation \eqref{lpdas}, either $m_{-}$ or $m_{+}$ is fixed and we get the section of the 2-dimensional histogram.
Therefore, the resultant histogram represents how many pairs of periodic orbits 
which have the determinate value $L_{\pm} = L_{\gamma} \pm L_{\gamma^{\prime}} \in [ m_{\pm} \Delta L_{\pm}, \, \left( m_{\pm} +1 \right) \Delta L_{\pm} ]$ 
are there in each divided domains $L_{\mp} = L_{\gamma} \mp L_{\gamma^{\prime}} \in [ m_{\mp} \Delta L_{\mp}, \, \left( m_{\mp} +1 \right) \Delta L_{\mp} ] $.
The lines to be slit up the joint distribution is perpendicular to the axis $L_{+} = L_{\gamma} + L_{\gamma^{\prime}}$, in other words, 
parallel to the axis $L_{-} = L_{\gamma} - L_{\gamma^{\prime}}$ ( see Fig.\ref{lpdascoordinate} (b) ).
Through this statistics, we will investigate the correlation among the orbits which are selected by the delta function.

We use the same coordinate system as used in the preceding density plots.
The distinction is the fact 
that the each distributions $C^{\pm} \left( L_{-}, L_{+} = \textrm{consts} \right)$ are taken along the direction for the difference direction $L_{-}$ 
with respect to each the fixed coordinates $L_{+} = L_{\gamma} + L_{\gamma^{\prime}}$.
As in the above density plot, the whole domain is segmentalised into $M_{-} \times M_{+} = 2^{12} \times 2^{12}$ cells.
Moreover, for looking effect of taking histogram width, we segmentalise the whole domain also into $M_{-} \times M_{+} = 2^{11} \times 2^{11} $ cells.
Then we can see the structures in slices more detail than in the density plot itself.
Also in this calculation, we take the histograms over the all periodic orbits with $n=17$ collisions 
and we takn the slices of the above joint distributions for each the radius of the four disks $r$, from $0.5$ to $1.0$ in the matter of 0.05.
\section{numerical analysis II : statistics of single periodic orbits}
\subsection{Length distribution for single periodic orbits $G_{\Delta L, m} \left( L \right)$}
%
%
As an assistant to understanding the preceding pair statistics $C^{\pm} \left( L_{-}, L_{+} \right)$, we need also to take some statistics for the single periodic orbits.
And we will ask the question whether behaviour of histogram can be varied according to manner of dividing its domain into several intervals.
Because of the brevity and the translucent correspondence to the foregoing distributions, we define the following distribution:
\begin{eqnarray}
G_{\Delta L, m} \left( L \right)
\equiv 
\sharp \{ \gamma \bigm| m \Delta L \leq  L_{\gamma}  < \left( m +1 \right) \Delta L \},
\\
\Delta L = \sup_{\gamma} \{ L_{\gamma} \} / M, \quad 
m = 0, 1,  \dots ,  M-1 \in \mathbf{N}.
\nonumber
\label{freqlen}
\end{eqnarray}
$\gamma$ is the index for a periodic orbit which includes repeated ones.
$M$ is the number of items which divide length domain $\left[ 0,\, \sup_{\gamma} \{ L_{\gamma} \}  \right]$, and  $m$ is the indices of each numbered intervals.
$\Delta L$ is the widths of each segmented intervals. 
This length histogram for single periodic orbits indicates how many periodic orbits are there in each intervals $\left[ m \Delta L,\, (m+1) \Delta L \right] $.
By use of this statistics, it can be seen how the histogram behaves according to the width $\Delta L$.
As long as there are no confusion in this paper, we abbreviate $G_{\Delta L, m} \left( L \right)$ as $G \left( L \right)$.

We calculated the histogram for the length among the single periodic orbits with $17$-collisions.
In this analysis, we adopted some dividing numbers $M$ for the above purpose.
\subsection{Cumulative number of length spectrum for single periodic orbits $N \left( L \right)$}
%
%
In the above statistics, there is bothering problem, how long widths do we adopt in those histograms.
Therefore, we need for the statistics which does not have to consider the width of the histograms.
One of the most simple statistics which does not take a form of histogram is a cumulative number density of length spectrum.

%
The cumulative number of length spectrum for the single periodic orbits is defined as how many periodic orbits which length is below $L$:
%
\begin{equation}
N \left( L \right) \equiv \sum_{\gamma=0}^{\infty} \theta \left( L - L_{\gamma} \right).
\label{defcumlensp}
\end{equation}
It is well known that this function has its asymptotic form for large $L$ \cite{PP1983},
\begin{equation}
N \left( L \right) \sim \exp \left( h_{top} L \right) / L \equiv N_{asympt} \left( L \right) ,
\label{asymptcumlenspdiv}
\end{equation}
where $h_{top}$ is topological entropy.

Our aim is not to take the asymptotic form but to reduce the deviation from its.
In the form of the cumulative function on its own, it is hard to recognise the fluctuation part as deviation from its asymptotic form 
because the number of $N \left( L \right)$ is more tremendous than that of $N_{fluct} \left( L \right)$ as increasing $L$.
As a result, we consider the following function:
\begin{equation}
N_{fluct} \left( L \right) = N \left( L \right) - N_{asympt} \left( L \right).
\label{fluctcumlenspdiv}
\end{equation}
That is, subtracting the asymptotic part $N_{asympt} \left( L \right)$ from the whole cumulative number density $N \left( L \right)$, we can get the fluctuating part $N_{fluct} \left( L \right)$.

Actual procedure is the following:
First, we took the cumulative number distribution $N \left( L \right)$ of the length spectrum.
The distribution expressed as $\exp \left( h_{top} L \right) / L$ in the asymptotic form.
Next, we performed the least-squares fitting with the function $N_{fit} \left( L \right) = A \exp \left( B L \right) / L$, where $A$ and $B$ are fitting parameters, in a reasonable local domain 
and we employ this fitted function as the asymptotic part $N_{asympt} \left( L \right)$ \eqref{asymptcumlenspdiv} of the cumulative function.
Finally, we subtract the fitted function $N_{fit} \left( L \right)$ from the whole distribution $N \left( L \right)$, 
and then we get the fluctuating part $N_{fluct} \left( L \right) $ \eqref{fluctcumlenspdiv} of the distribution.
%
%
%
\section{results}
\subsection{Distribution function of length difference / sum for periodic orbit pairs $C^{\pm}_{\Delta L_{\pm},m} \left( L_{\pm} \right)$}
\begin{figure}[htbp]
\begin{center}
\includegraphics[width=5.3cm,clip]{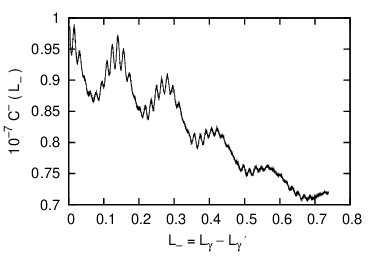}
\includegraphics[width=5.3cm,clip]{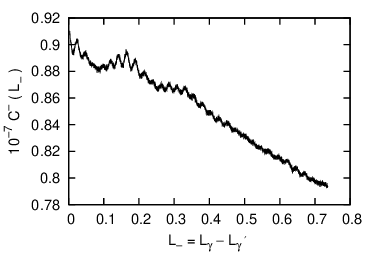}
\includegraphics[width=5.3cm,clip]{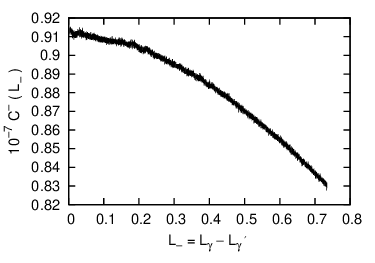}
\includegraphics[width=5.3cm,clip]{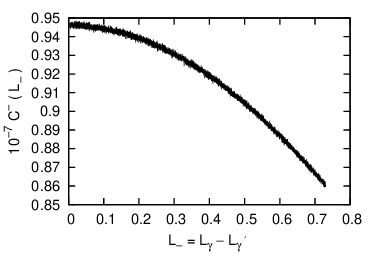}
\includegraphics[width=5.3cm,clip]{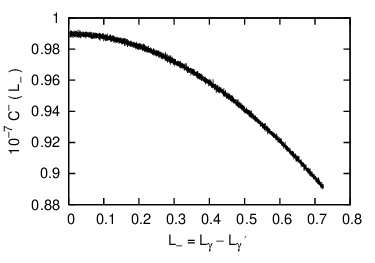}
\includegraphics[width=5.3cm,clip]{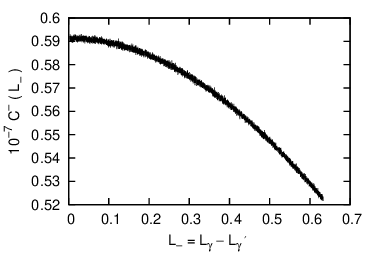}
\caption{{\bf Distribution of length difference for periodic orbit pairs $C^{-}_{\Delta L_{-},m} \left( L_{-} \right)$}: 
In the top, from left to right, $r=0.5,\,0.6,\,0.7$; in the bottom, from left to right, $r=0.8,\,0.9,\,1.0$}
\label{lpddist}
\vspace{5mm}
\includegraphics[width=5.3cm,clip]{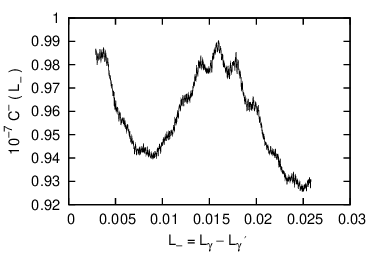}
\includegraphics[width=5.3cm,clip]{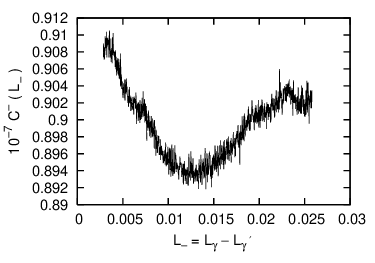}
\includegraphics[width=5.3cm,clip]{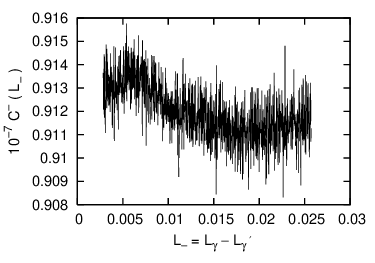}
\includegraphics[width=5.3cm,clip]{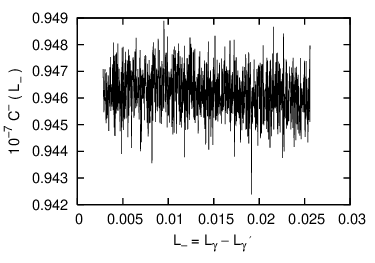}
\includegraphics[width=5.3cm,clip]{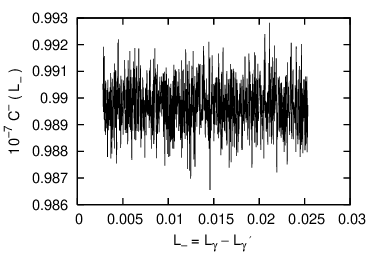}
\includegraphics[width=5.3cm,clip]{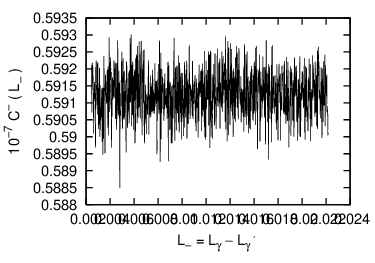}
\caption{{\bf Detailed distribution of length difference for periodic orbit pairs $C^{-}_{\Delta L_{-},m} \left( L_{-} \right)$} : 
in the top, from left to right, $r=0.5,\,0.6,\,0.7$; in the bottom, from left to right, $r=0.8,\,0.9,\,1.0$}
\label{dlpddist}
\end{center}
\end{figure}

\begin{figure}[htbp]
\begin{center}
\includegraphics[width=5.3cm,clip]{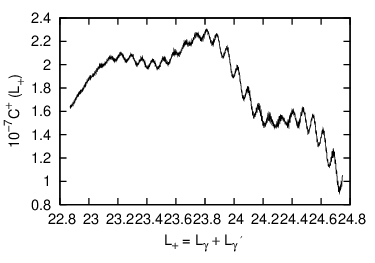}
\includegraphics[width=5.3cm,clip]{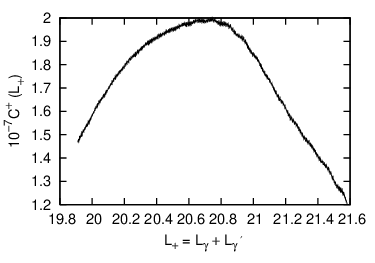}
\includegraphics[width=5.3cm,clip]{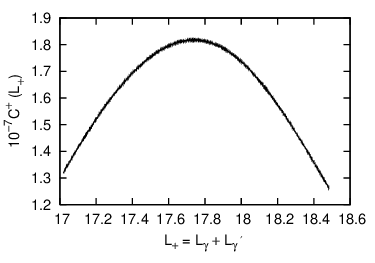}
\includegraphics[width=5.3cm,clip]{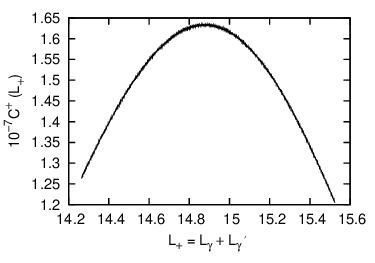}
\includegraphics[width=5.3cm,clip]{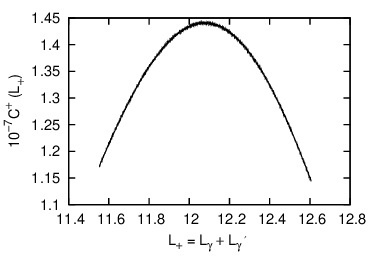}
\includegraphics[width=5.3cm,clip]{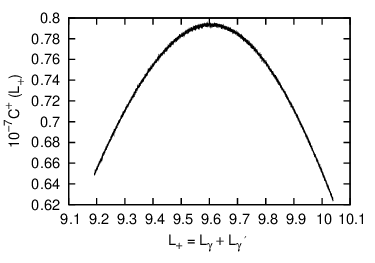}
\caption{{\bf Distribution of length sum for periodic orbit pairs $C^{+}_{\Delta L_{+},m} \left( L_{+} \right)$} : 
In the top, from left to right, $r=0.5,\,0.6,\,0.7$; in the bottom, from left to right, $r=0.8,\,0.9,\,1.0$}
\label{lpsdist}
\vspace{5mm}
\includegraphics[width=5.3cm,clip]{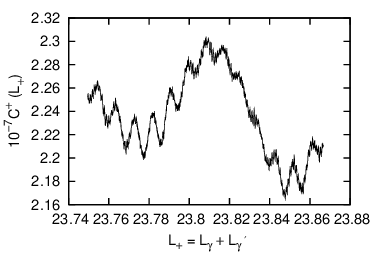}
\includegraphics[width=5.3cm,clip]{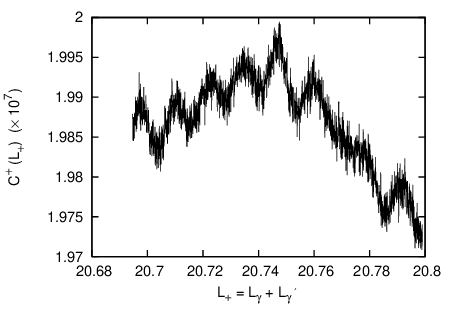}
\includegraphics[width=5.3cm,clip]{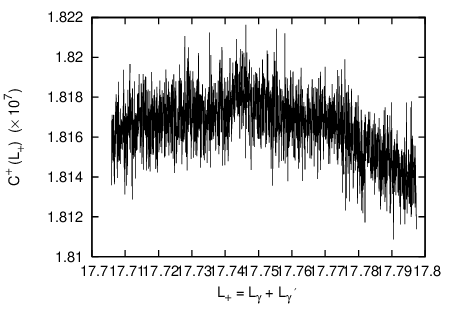}
\includegraphics[width=5.3cm,clip]{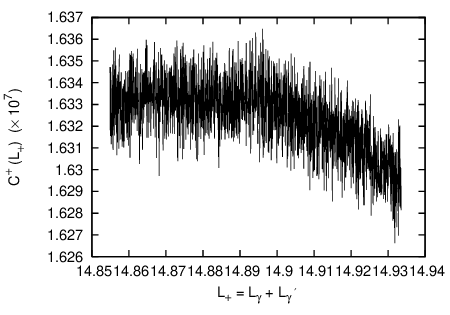}
\includegraphics[width=5.3cm,clip]{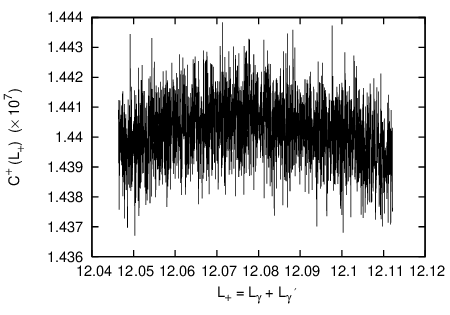}
\includegraphics[width=5.3cm,clip]{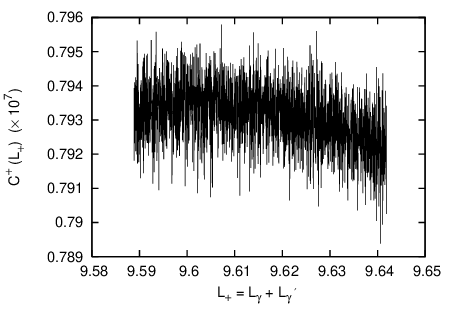}
\caption{{\bf Detailed distribution of length sum for periodic orbit pairs $C^{+}_{\Delta L_{+},m} \left( L_{+} \right)$} : 
In the top, from left to right, $r=0.5,\,0.6,\,0.7$; in the bottom, from left to right, $r=0.8,\,0.9,\,1.0$}
\label{dlpsdist}
\end{center}
\end{figure}

%
%
We took the distributions of length difference / sum for pairs $C^{-} \left( L_{-} \right)$ / $C^{+} \left( L_{+} \right)$ 
( Eq.\eqref{lpd} / Eq.\eqref{lps} ) among the periodic orbits with $n=17$ collisions for each radius $r$ ( Fig.\ref{lpddist} and Fig.\ref{lpsdist} ).
The domains $[0,\,\sup_{\gamma} \{ L_{\gamma} \pm L_{\gamma^{\prime}} \} ]$ are segmentalised into $M=2^{19}$ intervals.
And Fig.\ref{dlpddist} and Fig.\ref{dlpsdist} are detailed figures centring around the crest of the distributions Fig.\ref{lpddist} and Fig.\ref{lpsdist} respectively.
From the numerical analysis, the distributions of length difference / sum seem to depend on radius of disks at a glance ( Fig.\ref{lpddist} and Fig.\ref{lpsdist} ).
Beholding the length pairs' distributions from $r=0.5$ to $r=1.0$, we can find out variations with its radii in those distributions.
When $r=0.5$, it is easy to see the periodic peak structure, and furthermore, we recognise self-similar structures therein, 
While when $r = 1.0$, it is scarcely able to recognise such the structures.
But it is not clear whether such periodic peak structure and even more self-similar one of those distributions truly become disappeared as increasing $r$.
We need more detailed analysis to establish the facts of the matter.
%
%
%
%
%
%
%
%
\subsection{Joint distribution function of length difference \textit{and} sum for periodic orbit pairs $C^{\pm}_{\Delta L_{\pm},m_{-},m_{+}} \left( L_{-}, L_{+} \right)$}
\begin{figure}[htbp]
\begin{center}
\includegraphics[height=8cm]{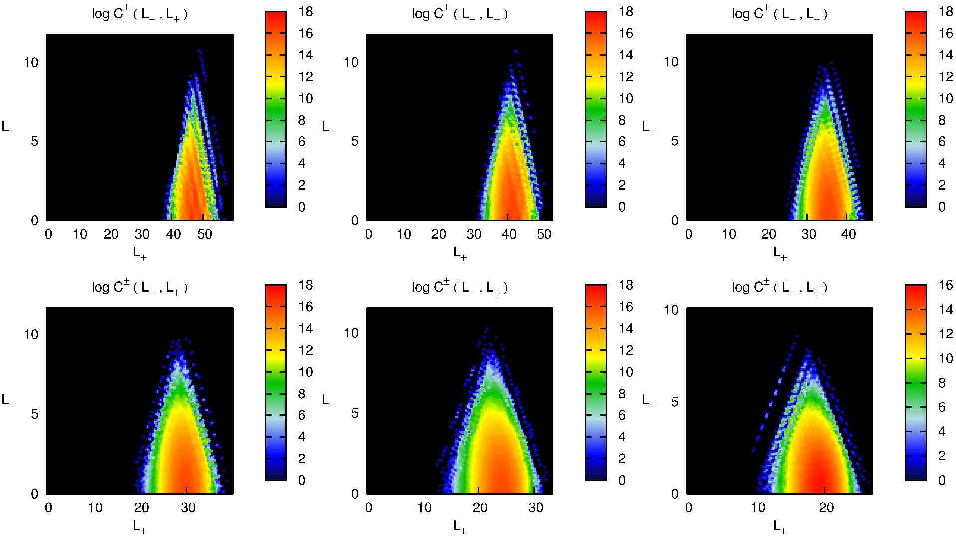}
\caption{(Colour Online). 
{\bf Density plots for joint distribution of length difference \textit{and} sum for periodic orbit pairs $C^{\pm}_{\Delta L_{\pm},m_{-},m_{+}} \left( L_{-}, L_{+} \right)$}:
In the top, from left to right, $r=0.5,\,0.6,\,0.7$; in the bottom, from left to right, $r=0.8,\,0.9,\,1.0$}
\label{lpdasdensityplot}
\vspace{4mm}
\includegraphics[height=8cm]{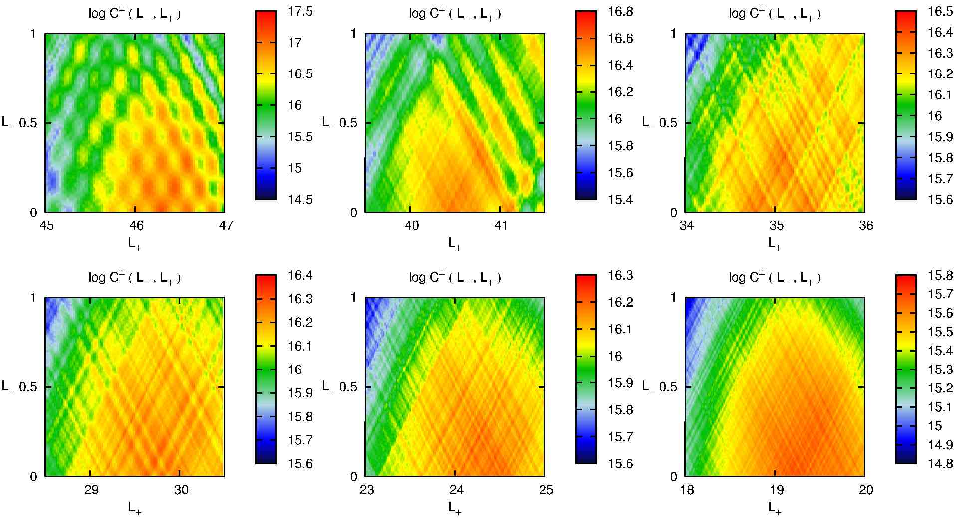}
\caption{(Colour Online). 
{\bf Detailed density plots for joint distribution of length difference \textit{and} sum for periodic orbit pairs $C^{\pm}_{\Delta L_{\pm},m_{-},m_{+}} \left( L_{-}, L_{+} \right)$}:
In the top, from left to right, $r=0.5,\,0.6,\,0.7$; in the bottom, from left to right, $r=0.8,\,0.9,\,1.0$}
\label{lpdasdensityplotdetail}
\end{center}
\end{figure}

%
%
Further analysis is to take the joint distribution of length difference \textit{and} sum 
for periodic orbit pairs $C^{\pm} \left( L_{-}, L_{+} \right)$ ( Eq.\eqref{lpdas} ).
As the above calculation, among the periodic orbits with $n=17$ collisions, we took the joint distributions for each radis $r$.
The each domains $[ 0, \, \sup_{\gamma} \{ L_{\gamma} - L_{\gamma^{\prime}} \} ] \times [ 0, \, \sup_{\gamma} \{ L_{\gamma} + L_{\gamma^{\prime}} \} ]$ 
are partitioned into $M_{-} \times M_{+} = 2^{12} \times 2^{12}$ cells.
Beholding the joint distributions from $r=0.5$ to $r=1.0$ ( Fig.\ref{lpdasdensityplot} ), we can find out variations with its radius in those distributions.
When $0.5 \leq r \leq 0.6$, rhombus pattern can be seen, while when $0.7 \leq r$, it is hard to recognise such structure at a glance.
But from the enlarged figures, the detail structure show up ( Fig.\ref{lpdasdensityplotdetail} ) out of relation to the values of radius $r$.
%
%
%
%
%
%
%
%
%
%
\subsection{Slice of joint distribution function of length difference \textit{and} sum $C^{\pm}_{\Delta L_{\pm},m_{-},m_{+}} \left( L_{-}, L_{+} \right)$}
\begin{figure}[htbp]
\begin{center}
\includegraphics[width=5.3cm]{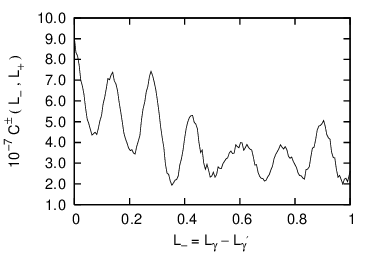}
\includegraphics[width=5.3cm]{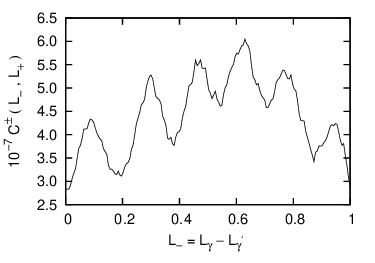}
\includegraphics[width=5.3cm]{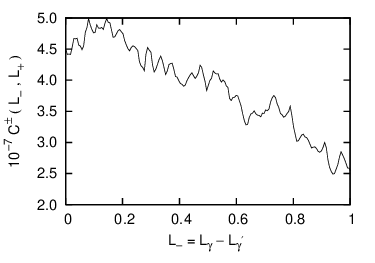}
\includegraphics[width=5.3cm]{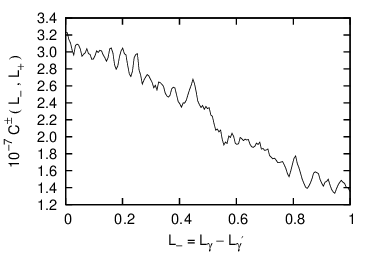}
\includegraphics[width=5.3cm]{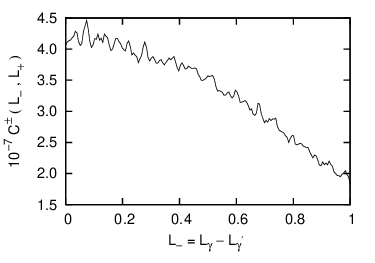}
\includegraphics[width=5.3cm]{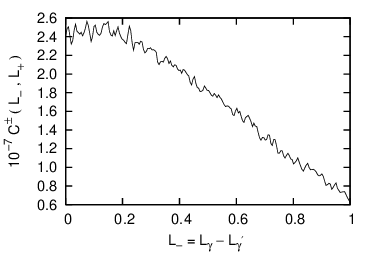}
\caption{{\bf Slices of density plots for joint distributions $C^{\pm}_{\Delta L_{\pm},m_{-},m_{+}} \left( L_{-}, L_{+} \right)$}:
the each domains are partitioned into $M_{-} \times M_{+} = 2^{11} \times 2^{11}$ cells;
In the top, from left to right, $r=0.5,\,0.6,\,0.7$; in the bottom, from left to right, $r=0.8,\,0.9,\,1.0$}
\label{lpdasdensityplotsliced11}
\vspace{7mm}
\includegraphics[width=5.3cm]{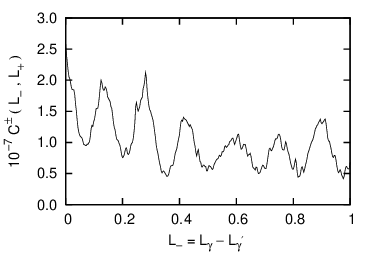}
\includegraphics[width=5.3cm]{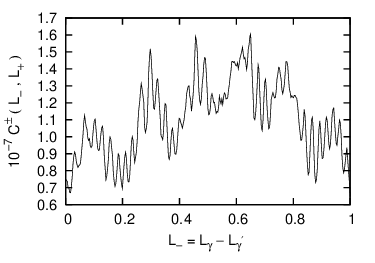}
\includegraphics[width=5.3cm]{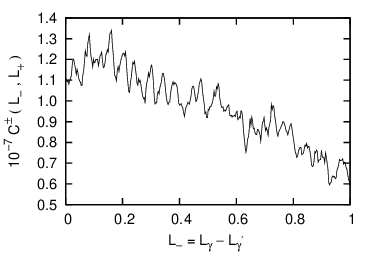}
\includegraphics[width=5.3cm]{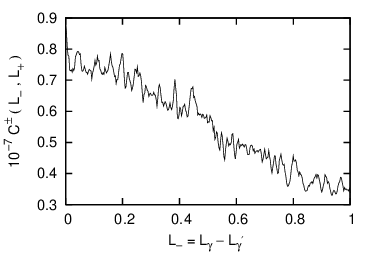}
\includegraphics[width=5.3cm]{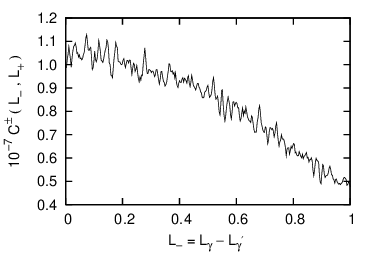}
\includegraphics[width=5.3cm]{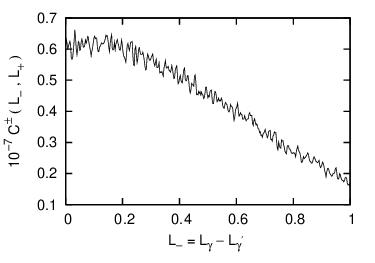}
\caption{{\bf Slices of density plots  for joint distribution $C^{\pm}_{\Delta L_{\pm},m_{-},m_{+}} \left( L_{-}, L_{+} \right)$}:
the each domains are partitioned into $M_{-} \times M_{+} = 2^{12} \times 2^{12}$ cells;
In the top, from left to right, $r=0.5,\,0.6,\,0.7$; in the bottom, from left to right, $r=0.8,\,0.9,\,1.0$}
\label{lpdasdensityplotsliced}
\end{center}
\end{figure}

%
%
In addition to the above analysis, we took the slice of the joint distributions along the lines where the sum of pair orbit's lengths is constant,
$L_{+} = L_{\gamma} + L_{\gamma^{\prime}} = \textrm{const.}$.
We took these histograms with no-logarithm plot for searching detail structure.
Needles to say, as in the preceding joint distributions, we referred to the periodic orbits with $n=17$ collisions for each radius $r$.
The each domains $[ 0, \, \sup_{\gamma} \{ L_{\gamma} - L_{\gamma^{\prime}} \} ] \times [ 0, \, \sup_{\gamma} \{ L_{\gamma} + L_{\gamma^{\prime}} \} ]$ 
are partitioned into $M_{-} \times M_{+} = 2^{12} \times 2^{12}, 2^{11} \times 2^{11}$ cells.
From the results of the slices of the joint distributions ( Fig.\ref{lpdasdensityplotsliced11} and Fig.\ref{lpdasdensityplotsliced} ), we can see the variations with its radii $r$.
When $0.5 \leq r \leq 0.6$, the periodic peak structure with broad periods can be seen, while when $0.7 \leq r$, it is hard to recognise such structure at a glance.
But in detailed structure,
slices of the joint distribution of length difference \textit{and} sum have the periodic peak structures in irrelevance to radius of disks $r$.
\subsection{Length distribution for single periodic orbits $G_{\Delta L, m} \left( L_{\gamma} \right)$}
\begin{figure}
\begin{center}
\hspace{0cm}
\subfigure[]{\includegraphics[width=8cm]{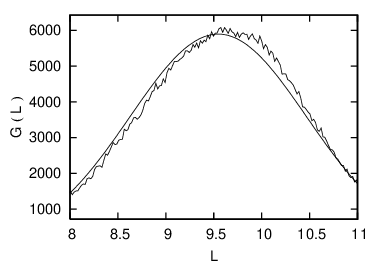}}
\hspace{0mm}
\subfigure[]{\includegraphics[width=8cm]{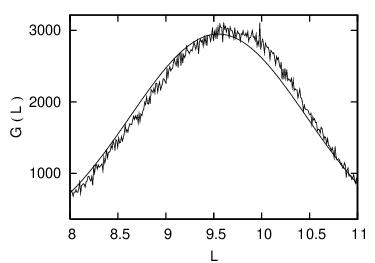}}
\caption{{\bf Length histograms for single periodic orbits $G_{\Delta L, m} \left( L \right)$:} the domain of the histograms is divided into (a) 700, (b)1400.}
\label{lengthhist}
\end{center}
\end{figure}

%
%
For more understanding to the foregoing pairs' distributions,
among the single periodic orbits with $17$ collision, we took the frequency distribution for their lengths ( Eq.\eqref{freqlen} ) for the case of $r=1.0$.
And we adopt the numbers $M$ of all intervals, $M=700, 1400$, to see effect by width of the histogram.
Then we can see the periodic peak structures in these histograms ( Fig.\ref{lengthhist} ) 
and there is significant deviation from the corresponding gaussian distribution.
The periodic peak structure become to be split as varying the number of intervals.
But rough structures of them remain, which is indication that there must be some structure in essential distribution independently of the width of the histogram.
\subsection{Cumulative number of length spectrum for single periodic orbits $N \left( L \right)$}
\begin{figure}
\begin{center}
\hspace{0cm}
\subfigure[]{\includegraphics[width=8cm]{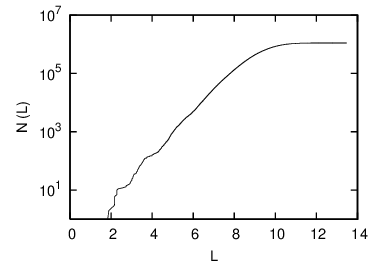}}
\hspace{0mm}
\subfigure[]{\includegraphics[width=8cm]{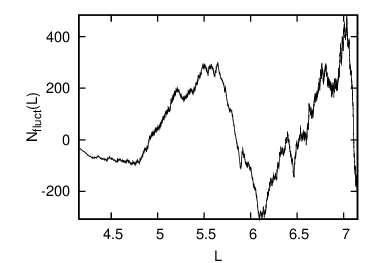}}
\caption{{\bf Cumulative number of length spectrum:} (a) numerical result for cumulative number $N (L)$, (b) fluctuating part $N_{flct}(L)$}
\label{cumlspect}
\end{center}
\end{figure}
%
%
Because we would look at the problem of the manner of taking width of histogram from the outside,
we analysed the cumulative number density $N \left( L \right)$ of the length spectrum \eqref{defcumlensp} in the case of $r=1.0$.
In doing so, we referred to the periodic orbits which has $2 \leq n \leq 17$ collisions.
After the least-squares fitting, we subtracted the fitted function $N_{fit} \left( L \right)$ from the whole cumulative number $N \left( L \right)$
and we get the fluctuating part $N_{fluct} \left( L \right)$ of the distribution \eqref{fluctcumlenspdiv}.
Fig.\ref{cumlspect} (b) shows the fluctuating part $N_{fluct} \left( L \right)$ of the cumulative number density of the length spectrum.
There is exactly the periodic peak structure and moreover seemed to be self-similar structure.
\section{summary and discussion}
\subsection{Summary}
%
%
Let us summarise the main points that have been mentioned above.
Clarification of periodic orbits' correlation is indispensable for semiclassical reasoning of spectral fluctuation.
In the 4-disk billiard system, we performed the statistical analysis for the pairs among periodic orbits and for the single ones.
From the viewpoint of brevity, we presented the statistical analysis on the differences and the sums which appear in the semiclassical spectral form factor:
for the periodic orbit pairs, both the distribution function of length difference / sum  $C^{-} \left( L_{-} \right)$ / $C^{+} \left( L_{+} \right)$ 
and the joint distribution function of length difference \textit{and} sum $C^{\pm} \left( L_{-}, L_{+} \right)$
and for the single periodic orbits, 
both the length distribution $G \left( L \right)$ 
and the cumulative number of length spectrum $N \left( L \right)$.
In those analysis, we found some universality of deviation from relevant Gaussian distribution, concretely saying, the periodic peak structure.
\subsection{Distribution function of length difference / sum for periodic orbit pairs $C^{\pm}_{\Delta L_{\pm},m} \left( L_{\pm} \right)$}
%
%
As the first step, we took the distribution of length difference / sum for periodic orbit pairs $C^{-} \left( L_{-} \right)$ / $C^{+} \left( L_{+} \right)$ 
( Figs.\ref{lpddist},\ref{dlpddist} / Figs.\ref{lpsdist},\ref{dlpsdist} ).
From the consistency with the analysis made by Shudo and Ikeda, it was anticipated that there must be periodic peak structure in those distributions.
These histograms were calculated for each the radius of the disks $r$, from $0.5$ to $1.0$ in the matter of $0.05$.
In brief tendency, the histograms present gaussian distribution.
Focusing attention on the detail structures,
For small $r$ ( $0.5 \leq r \leq 0.7$ ), there seemed to be the periodic peak structure and they present self-similar ones, 
while for larger $r$ ( $0.8 \leq r \leq 1.0$ ), harder we recognize such structures.
But we have no assurance that the periodic peaks structure are certainly disappeared in the large $r$.
\subsection{Joint distribution function of length difference \textit{and} sum for periodic orbit pairs $C^{\pm}_{\Delta L_{\pm},m_{-},m_{+}} \left( L_{-}, L_{+} \right)$}
%
%
Inspired from the configuration of the semiclassical form factor \eqref{sffscl}, 
we analysed the joint distributions of length difference \textit{and} sum for the periodic orbit pairs \eqref{lpdas}.
These joint distributions are taken in the form of the density plots ( Figs.\ref{lpdasdensityplot},\ref{lpdasdensityplotdetail} ).
As is the case with the above analysis, 
these density plots were taken for each the radius of the disks $r$, from $0.5$ to $1.0$ in the matter of $0.05$.
Focusing attention on the detail structures,
while for small $r$ ( $0.5 \leq r \leq 0.7$ ), there seemed to be the periodic peak structure and they present self-similar ones, 
Although for large $r$ ( $0.8 \leq r \leq 1.0$ ) in the whole density plots, we can hardly recognise such as the above structures.
But from the detail density plots, evidently, we become aware of the periodic peak structure.
\subsection{Slice of joint distribution function of length difference \textit{and} sum $C^{\pm}_{\Delta L_{\pm},m_{-},m_{+}} \left( L_{-}, L_{+} \right)$}
%
%
For intelligibleness, we taken the slices of the joint distributions of length difference \textit{and} sum, i.e. the sections of the above density plots 
( Figs.\ref{lpdasdensityplotsliced11},\ref{lpdasdensityplotsliced} ).
And we taken these distributions in anti-logarithm for vertical axis of the histogram.
Also in these sections, we can recognise that the periodic peak structures remain as varying $r$.
Therefore, we have further confirmation that there is some universality among the above density plots.
However, the manner of taking histogram's width is left open problem.
\subsection{Length distribution for single periodic orbits $G_{\Delta L, m} \left( L_{\gamma} \right)$}
%
%
For the understanding for the periodic peak structures as observed in the above analysis,
we took the frequency distribution $G \left( L \right)$ for the length of the orbits which has $n=17$ strike points ( Fig.\ref{lengthhist} ).
This analysis performed in the value of $r=1.0$. 
Indeed, the periodic peak structure can be seen in this analysis.
But the spread of the peak seemed to be varied according to how we take the widths of the histograms.
If the periodic peak structure has self-similar structure, it is critical matter how to take the histogram width.
\subsection{Cumulative number of length spectrum for single periodic orbits $N \left( L \right)$}
%
%
To avoid the bothering `histogram width' problem in the anterior analysis,
there is need for the analysis which is free from the histogram.
We took the cumulative number for the length spectrum in $r=1.0$.
From it, we ultimately reduced the fluctuating part $N_{fluct} \left( L \right)$ through the least-square fitting.
In the fluctuating part, we observed the peak structure and moreover recognised self-similar structure.
Therefore, we assure there are indeed the periodic peak structure of the length histogram and besides that of the joint distributions.
\subsection{Picture of periodic peak structure}
%
%
From the above-mentioned analysis, we can claim as the following:
In the joint distribution of length difference \textit{and} sum $C^{\pm} \left( L_{-}, L_{+} \right)$, there are in fact the periodic peak structure in a large, 
while in the histograms of length difference / sum $C^{-} \left( L_{-} \right)$ / $C^{+} \left( L_{+} \right)$, we cannot say that the periodic peak structure exists irrespective of $r$.

As the above mentioned, the periodic peak structure of the joint distribution has pruning-proof property.
If such structure contribute to the off-diagonal part of the semiclassical form factor, 
the off-diagonal part has also pruning-proof property.
This means the following:
The correlation is the constraint condition for which preserve the periodic peak structure,
However we do not know yet how many pairs of periodic orbits contribute to the periodic peak structure, 
and whether the number of periodic orbits which contribute to the form factor can be varied as success of pruning. 
Stated another way, we do not know more of the fact that the pruning fronts \cite{Cvitanovic1991} is a constraint condition for the correlation.

We consider the fact that there is seemingly contradictory behaviour 
between the distribution of length difference / sum and the joint distribution of length difference \textit{and} sum.
From the density plots ( Fig.\ref{lpdasdensityplotdetail} ), 
it appears that the peak lines run along a oblique direction against the lines, $L_{\gamma} - L_{\gamma^{\prime}} = \textit{const.}$ and $L_{\gamma} + L_{\gamma^{\prime}} = \textit{const.}$.
Furthermore, there is an apparent relation between the distribution $C^{-} \left( L_{-} \right)$ / $C^{+} \left( L_{+} \right)$ 
and the joint distribution $C^{\pm} \left( L_{-}, L_{+} \right)$ as follows:
\begin{eqnarray}
C^{-}_{\Delta L_{-},m_{-}} \left( L_{-} \right) &=& \sum_{m_{+}=0}^{M_{+}-1} C^{\pm}_{\Delta L_{\pm},m_{-},m_{+}} \left( L_{-},\, L_{+} = m_{+} \Delta L_{+} \right),
\\
C^{+}_{\Delta L_{+},m_{+}} \left( L_{+} \right) &=& \sum_{m_{-}=0}^{M_{-}-1} C^{\pm}_{\Delta L_{\pm},m_{-},m_{+}} \left( L_{-} = m_{-} \Delta L_{-},\, L_{+} \right).
\label{slpd}
\end{eqnarray}
Consequently, the periodic peak structure in the joint distribution for the periodic orbit pairs seem to be cancelled 
when they are summed up to the distribution of length difference / sum.
From the perspective of the semclassical form factor \eqref{sffscl}, 
it is not so important to discuss absence or presence of the peak structure in the distribution of length difference / sum.

In the second place, we consider the cause of the rhombus pattern in the density plots for the joint distributions of length difference \textit{and} sum.
And so let us review the coordinate system again ( see Fig.\ref{lpdascoordinate} (a) ).
The rhombus pattern are formed along the lines $L_{\gamma} - L_{\gamma^{\prime}}$ and $L_{\gamma} + L_{\gamma^{\prime}}$.
This fact indicates that the correlation ( periodic peak structure ) exists originally in the length histogram for single periodic orbits.

From the last analysis, we got the periodic peak structure from the statistics which does not assume the form of a histogram.
Therefore, we can claim that the preceding length distribution for single periodic orbits which is shown as a histograms have indeed the periodic peak structure. 
This indicates that such deviation from asymptotic behaviour $\exp \left( h_{top} L \right) / L$ must contribute to the off-diagonal term of the semiclassical form factor.
The statistics which we have analysed has probably self-similar structure,
then we must treat the class interval of the histogram in a prudent manner.

Toward further understanding, we must make the correlation's picture more detail.
It is natural to question whether the periodic orbit pairs which contribute to the periodic peak structure can be changed as varying $r$.
Owing to structural stability of the periodic peak structure,
all we can say is that the combination of the periodic orbits which constructs particular peak among periodic peak structure must be fixed.
For this reason, such contributing orbits must be structurally stable.
Against the interpretation mentioned at preceding sentences, structural unstable pairs, which consist of bifurcated orbits ( e.g. Fig.\ref{1stpruning} and Fig.\ref{pruningiebifurcation} ),
after all, do not contribute to particular peak. 
Because compared to length difference of structural stable ones, those of structurally unstable ones are variable and ultimately come to vanish as increasing $r$.

Next, we would like to mention a manner of sorting out the contributions to the periodic peak structure:
which kind of set must we classify the contribution to such structure according to, a set of single periodic orbits or a set of pairs of periodic orbit ?
From the histogram of length for single periodic orbits and the preceding considerations for it, 
we speculate that assemblage to be performed is to censor in light of a set of single periodic orbits.

One of special properties of billiard system is the inseparably relation between action and period \eqref{periodactionforbilliard}.
Therefore, at a glance, it seems to be reasonable that the above statistical properties are highly special.
However, essentially for a general system, action and period are closely related to each other.
\begin{equation}
T \left( E \right) = \frac{d S \left( E \right)}{d E} .
\label{periodaction}
\end{equation}
Consequently, the preceding idea for the correlation, i.e. the periodic peak structure, must be extended.
\subsection{Conclusion}
We have performed numerical calculation of not non-periodic orbits but actual periodic orbits in the 4-disk billiard system.
And we analysed the distributions of length difference \textit{and} / \textit{or} sum among the periodic orbit pairs 
and the histogram and the cumulative function of length spectrum for the single periodic orbits.
From these analysis, it is clear that 
the periodic peak structures are observed in the joint distribution of length difference \textit{and} sum, 
that is, the correlation to be required is observed in the statistics for the quantities which the semiclassical form factor explicitly consists of.
And this is ultimately a result from the correlation which the length spectrum for single periodic orbits has.
And notably such property remains unaffected by pruning.
Furthermore, we ask whether pruning or bifurcation can become constraint condition for the correlation, 
and we conclude if either of the pair of orbits come into bifurcation then such the pair cannot contribute to the periodic peak structure.
That is, if the periodic peak structure has very significant contribution to the form factor, we do not have to care bifurcation.
\section*{Acknowledgment}
Author thanks to Tetsuro Konishi because I often have a fruitful discussion with him
and he read this manuscript in detail and gave me many adequate comments.
Author also thanks to Akira Shudo for a great deal of valuable advice mainly on periodic orbit correlation.
Yutaka Ishii and Takehiko Morita for introducing me to billiard problem from the aspect of dynamical systems.
Mitsusada M. Sano gave me ingenuous comments about this research.
Takuhisa Harayama gave me many pieces of valuable advice on numerical analysis of billiards.
Author also thanks to referees for a nice comments to improve this paper.
This research is partially supported by the grant-aid for the Nagoya University 21st Century COE ( Centre Of Excellence ) Program ``ORIUM (The Origin of the Universe and Matter)''.
%
%
\section*{References}
\bibliographystyle{apsrev}
\newcommand{\noopsort}[1]{}

\end{document}